\documentclass[A_4paper,12pt, final]{article}
\pdfoutput=1

\usepackage{jheppub}

\usepackage[utf8]{inputenc}
\usepackage{graphicx}
\usepackage{hyperref}
\usepackage{amsmath}
\usepackage{color}
\usepackage{epstopdf}
\usepackage{float}
\usepackage{upgreek}
\usepackage{amssymb}
\usepackage{pdflscape}
\usepackage[nice]{nicefrac}
\usepackage[dvipsnames]{xcolor}
\usepackage{subcaption,graphicx}
\usepackage[numbers, sort&compress]{natbib}
\usepackage{braket}
\usepackage{comment}
\usepackage{slashed}

\newcommand{\rhoR}{\rho_R}

\usepackage{xspace}
\newcommand*{\fig}{fig.\@\xspace}
\newcommand*{\eq}{eq.\@\xspace}
\newcommand*{\cf}{cf.\@\xspace}
\newcommand*{\eqs}{eqs.\@\xspace}
\newcommand*{\rhs}{r.h.s.\@\xspace}

\newcommand{\tp}{\braket{\varphi}}

\title{On Sphaleron Heating in the Presence of Fermions}

\author{Marco Drewes,}
\author{Sebastian Zell}

\affiliation{Centre for Cosmology, Particle Physics and Phenomenology -- CP3,
	Universit\'e catholique de Louvain, B-1348 Louvain-la-Neuve, Belgium}
\emailAdd{marco.drewes@uclouvain.be}
\emailAdd{sebastian.zell@uclouvain.be}

\abstract{Axion-like particles with a coupling to non-Abelian gauge fields at finite temperature can experience dissipation due to sphaleron heating. This could play an important role for warm inflation or dynamical dark energy. We investigate to what degree the efficiency of this non-perturbative mechanism depends on the details of the underlying particle physics model. For a wide range of scenarios and energy scales, we find that a previously discussed suppression of sphaleron heating by light fermions can be alleviated. As an outlook, we point out that fermionic effects may provide a new mechanism for ending warm inflation.}

\date{}

\makeatletter
\gdef\@fpheader{\phantom{text}}
\makeatother

\begin{document}

\maketitle

\section{Introduction}

The evolution of scalar and pseudo-scalar fields in a hot plasma can potentially shape the cosmos in various ways,
including cosmic inflation,  phase transitions, moduli,  Dark Matter, vacuum decay and the stability of the Standard Model (SM) of particle physics. 
In the present work, we are specifically concerned with the evolution of an axion or axion-like field\footnote{In the following, we refer to all pseudoscalar fields with a potential that exhibits an approximate shift symmetry and an axion-like coupling to gauge fields as axions without distinguishing the QCD axion from axion-like particles.} 
that couples to a hot plasma of non-Abelian gauge fields via the operator 
\begin{equation}\label{AxionCoupling}
\frac{\varphi}{f}
\frac{g^2 \text{Tr}\, G_{\mu\nu} \tilde{G}^{\mu\nu}}{16 \pi^2} \equiv
 \frac{\varphi}{f} q \;,
\end{equation}
where $G_{\mu\nu}$ is the field-strength tensor, $\tilde{G}^{\mu\nu} = \frac{1}{2}\epsilon^{\mu\nu\rho\sigma}G_{\rho\sigma}$, and $f$ the axion decay constant.  
At a sufficiently high temperature $T$, the interaction \eqref{AxionCoupling} leads to $\varphi$-dissipation due to an effect known as \emph{sphaleron heating}.  
Qualitatively this may be understood by noticing that a non-zero time derivative $\dot{\varphi}$ induces an effective force in the space of gauge field configurations which leads to an overall gain in their kinetic energy, i.e., a net transfer of energy from $\varphi$ to the plasma
(cf.~\cite{Arnold:1987zg,Trodden:1998ym,Mirbabayi:2022cbt} for an intuitive analogue in classical mechanics).
When coupled to a pure Yang-Mills theory, the dissipation coefficient $\gamma_{\text{s}}$ for $\varphi$ is proportional to the sphaleron rate $\Gamma_{\text{sph}}$, which is unsuppressed at high temperatures \cite{Kuzmin:1985mm,Arnold:1996dy} (see \cite{Moore:2010jd,DOnofrio:2014rug,Laine:2022ytc} and references therein for more recent discussions). 
However, it has already been pointed out in \cite{McLerran:1990de} and was more recently discussed in \cite{Berghaus:2019whh,Berghaus:2020ekh} that the presence of light fermions can strongly suppress $\gamma_\text{{s}}$. 
Since all known elementary fermions were massless at temperatures above roughly 160 GeV \cite{DOnofrio:2014rug}, this may tempt one into the conclusion that sphaleron heating is not relevant in the real world. In the present work, we address the issue how generic the suppression of sphaleron heating due to gauge interactions with light fermions is.

While axions are probably mostly studied in the context of the strong CP-problem \cite{Peccei:1977ur,Peccei:1977hh,Wilczek:1977pj,Weinberg:1977ma}\footnote{There are indications that eternal de Sitter states must not exist in a theory of quantum gravity due to fundamental inconsistency caused by quantum breaking \cite{Dvali:2013eja,Dvali:2014gua,Dvali:2017eba,Dvali:2018fqu,Dvali:2018jhn,Dvali:2020etd,Dvali:2021kxt} (see \cite{Obied:2018sgi,Andriot:2018wzk,Garg:2018reu,Ooguri:2018wrx} for analogous conjectures in string theory). If this is the case, the presence of a QCD axion becomes a consistency requirement, independently of naturalness arguments \cite{Dvali:2018dce,Dvali:2022fdv}. 
  On the other hand, the existence of a strong CP-problem has recently been put into question due to the realisation that the phases from the mass terms and from the instanton effects are aligned if one reverses the order of the infinite volume limit and the sum over topological sectors \cite{Ai:2020ptm}. That is, if these limits are taken the other way around, there is no physical phase, and hence no CP-problem \cite{Ai:2022htq}.}
(c.f.~e.g.~\cite{DiLuzio:2020wdo} for references)
and in their role as Dark Matter candidates \cite{Preskill:1982cy,Dine:1982ah,Abbott:1982af} (c.f.~e.g.~\cite{Adams:2022pbo} for references),
they can also be motivated in the context of inflation \cite{Freese:1990rb}. 
In particular, axions can enable a consistent realization of \emph{warm inflation} \cite{Berera:1995ie}. In this alternative to standard (cold) inflation, dissipation due to the presence of a thermal bath during inflation plays a key role in maintaining the slow roll conditions,\footnote{Here and throughout this work we refer to the scenario where dissipation dominates over Hubble damping as warm inflation, i.e., the strong regime in the language of \cite{Laine:2021ego}.}
 thermal fluctuations contribute to the generation of cosmic perturbations,\footnote{There have been various phenomenological studies on perturbations generated during warm inflation see \cite{Berera:2008ar,Rangarajan:2018tte,Kamali:2023lzq,Ballesteros:2023dno,Montefalcone:2023pvh} for (partial) reviews. Characteristic observable predictions have been reported, including a suppression of the tensor-to-scalar ratio \cite{Graham:2009,Bastero-Gil:2011rva, LopezNacir:2011kk} and sizable non-Gaussianities with distinct shapes \cite{Gupta:2002kn,Gupta:2005nh,Moss:2007cv,Moss:2011qc,Bastero-Gil:2014raa,Mirbabayi:2022cbt}. 
Even when thermal effect do not dominate the background evolution, they can have a significant influence on the spectrum of perturbations (see e.g., \cite{Mirbabayi:2022cbt}). 
As discussed subsequently, however, important issues can already arise at the level of the background evolution \cite{Yokoyama:1998ju}.
Correspondingly, we shall not study perturbations in the present work, but only investigate the feasibility of the sphaleron heating mechanism at the background level.}
and the thermal and expansion histories of the universe are altered insofar as that there is no separate reheating epoch.\footnote{
The thermal history can in principle be probed due to the thermal emission of gravitational waves \cite{Ghiglieri:2015nfa}, which is known in the SM  \cite{Ghiglieri:2015nfa,Ghiglieri:2020mhm} and some extensions \cite{Ringwald:2020ist,Ringwald:2022xif}, including  axion inflation \cite{Klose:2022knn,Klose:2022rxh}.
Moreover, in standard inflation the reheating epoch leaves an imprint in the CMB \cite{Liddle:2003as,Martin:2010kz,Adshead:2010mc}
and can depend on microphysical aspects \cite{Drewes:2015coa,Drewes:2017fmn,Drewes:2019rxn}.
The sensitivity of next generation CMB observatories like LiteBIRD \cite{Sugai:2020pjw} and CMB-S4 \cite{CMB-S4:2020lpa}
can be sufficient to measure the inflaton coupling and reheating temperature \cite{Drewes:2019rxn,Drewes:2022nhu,Drewes:2023bbs}. A systematic exploration of the perspectives to probe warm inflation in along these lines would be desirable.
} 
While this idea was proposed long ago \cite{Berera:1995ie}, a challenge for implementations at the microphysical level arises due to the problem that interactions which generate a sufficiently large dissipation coefficient tend to modify the effective potential in a way that spoils its flatness 
\cite{Yokoyama:1998ju}.\footnote{This can be understood from the well-known fact that perturbative contributions to the dissipation rate and the finite temperature effective potential come from the real and imaginary parts of the same diagrams, c.f.~\cite{Buldgen:2019dus,Kainulainen:2021eki} and references therein.}
Warm axion inflation \cite{Mohanty:2008ab,Mishra:2011vh,Visinelli:2011jy,Ferreira:2017lnd,Ferreira:2017wlx,Kamali:2019ppi,Berghaus:2019whh} can avoid this issue since the shift symmetry of the axion ensures that the flatness of the potential is not spoiled by quantum corrections. A concrete model named \emph{minimal warm inflation} \cite{Berghaus:2019whh} implements axion inflation driven by sphaleron heating.

Finally, it has been pointed out that sphaleron damping in a dark sector can make a dynamical contribution to dark energy \cite{DallAgata:2019yrr,Berghaus:2020ekh} (see also \cite{Poulin:2018cxd,Berghaus:2019cls,Berghaus:2023ypi}) and moreover influence the production of axionic Dark Matter \cite{Papageorgiou:2022prc}. The goal of the present work is to show how a possible suppression of these mechanisms by light fermions can be alleviated. We leave a detailed study of concrete scenarios and their phenomenology for future work.

\section{Sphalerons and axion dynamics at finite temperature and chemical potential}

Let us consider a situation in which $\varphi$ couples to a QCD-like gauge theory with $N_c$ colours and a gauge coupling $g$:
\begin{equation} \label{originalModel}
    \mathcal{L} = -\frac{1}{2} \text{Tr}\, G_{\mu\nu} G^{\mu\nu} +\left(\frac{i}{2} \bar{\psi} \slashed{D} \psi + \text{h.c.}\right) -\frac{1}{2} \partial_\mu \varphi \partial^\mu \varphi + \frac{\varphi}{f} q - V(\varphi) \;.
\end{equation}
Here $V(\varphi)$ is the axion potential, $D_\mu = \partial_\mu - i g G_\mu$ denotes the covariant derivative of fermions, $G_\mu$ is the non-Abelian gauge field, we use the metric signature $\eta_{\mu\nu}=(-1,+1,+1,+1)$, and $q$ is defined in \eq \eqref{AxionCoupling}.
One can estimate in the high temperature regime \cite{Arnold:1996dy,Moore:2010jd}
\begin{equation} \label{GammaSph}
    \Gamma_{\text{sph}}= c_{\text{sph}} (\alpha N_c)^5 T^4  \;,
\end{equation}
where $\alpha = g^2/(4\pi)$ and $c_{\text{sph}}$ is a numerical factor of order $1$.

The Lagrangian \eqref{originalModel} contains only the axion and gauge fields, while matter in the real world is made of fermions. 
Though the existence of a dark sector that is free of any matter is in principle conceivable, all fermions in the universe being singlet under the gauge interaction would represent a rather special case. 
Therefore, it is important to determine how the addition of fermions alters the above theory.
Following \cite{McLerran:1990de}, we for simplicity assume that there is only one family of massless Dirac fermions $\psi$ that transform under the fundamental representation of the gauge group. 
At the classical level, there are two conserved fermionic currents, which can be written as $\bar{\psi}\gamma^0\psi$ and $\bar{\psi}\gamma^0\gamma^5\psi$, or alternatively as $\bar{\psi}_R\gamma^0\psi_R$ and $\bar{\psi}_L\gamma^0\psi_L$, where $\psi_{R,L} = P_{R,L} \psi$ and $P_{R,L} = (1\pm\gamma_5)/2$ are the usual chiral projectors.
Due to the Adler–Bell–Jackiw anomaly \cite{Adler:1969gk,Bell:1969ts}, the only fermion number density which is exactly conserved is 
$\bar{\psi}_R\gamma^0\psi_R + \bar{\psi}_L\gamma^0\psi_L$,\footnote{\label{ThermalMassFootnote}Note that thermal masses generated by the gauge interactions \cite{Klimov:1981ka,Weldon:1982bn} do not flip the chirality.} 
while the combination 
$n_\psi \equiv \bar{\psi}_R\gamma^0\psi_R - \bar{\psi}_L\gamma^0\psi_L$ 
is violated at a rate $\sim \Gamma_{\text{sph}}/T^3 \propto g^{10} T$.
Comparing this to typical damping rates in the plasma, which are of order $\sim g^4 T$ to $\sim g^2 T$ (depending on momentum and scattering angle \cite{Arnold:2002zm,Ghiglieri:2020dpq}), one observes a separation of timescales for $g\ll 1$, which justifies treating $n_\psi$ as approximately conserved and defining a slowly varying chemical potential $\mu_\psi$, which in the regime $\mu_\psi\ll T$  is linearly related to the charge $\mu_{\psi}\simeq 6 n_{\psi}/T^2$.\footnote{Alternatively, we may think of $\psi$ as a combination $\psi = \psi_R + \psi_L$ of two Weyl fermions $\psi_{R,L}$, in which case $\mu_{R,L}\simeq 6 n_{R,L}/T^2$ are simply the chemical potentials of their matter-antimatter asymmetries and $\mu_\psi = \mu_R-\mu_L$. The arguments in the following would in fact also apply if only one of the Weyl fermions were present.}

\subsection{Mechanism for disabling sphaleron heating}\label{Sec:Disabeling}

In a thermal state, the equations of motion for the axion and fermionic chemical potential read \cite{McLerran:1990de}:\footnote{We do not strictly distinguish between the tree-level potential $V$ and the effective potential for $\tp$ during inflation here because the modifications of the latter are small in minimal warm inflation.} 
\begin{align}
-\ddot{\tp} &   = V' +  \left(3H + \gamma_{\text{s}}\right)  \dot{\tp} + \frac{\gamma_{\text{B}} T^2}{6 f} \mu_\psi\;,  \label{axionEOMFullHubble}\\
\dot{\mu}_\psi &=  -3H\mu_\psi - \frac{6f}{T^2}\left(\gamma_{\text{s}}  \dot{\tp} + \frac{\gamma_{\text{B}} T^2}{6 f} \mu_\psi\right)  \label{fermionEOMFullHubble} \;,
\end{align}
where prime denotes derivative with respect to time and brackets $\braket{\ldots}$ indicate thermal and quantum averages.
In general, the dissipation coefficient $\gamma_{\text{s}}$ that is central to warm inflation or dark energy can be obtained by evaluating the thermal expectation value of retarded correlators at some frequency $\omega \sim \dot{\varphi}/\varphi$, cf.~\cite{Buldgen:2019dus} and references therein. 
For the specific case of the operator \eqref{AxionCoupling}, different regimes can be distinguished by comparing $\omega$ to $T$, cf.~section 5 in \cite{Laine:2021ego}.
For $\omega \ll \alpha^2 T$, the various perturbative contributions \cite{Laine:2011xm} are dominated by the effect of the non-perturbative sphaleron heating effect and governed by the rate \eqref{GammaSph},
\begin{equation} \label{sphaleronBackreaction}
    - \left\langle q  \right\rangle  =  \frac{\Gamma_{\text{sph}}}{T}\left(\frac{ \dot{\tp}}{f} + 2 \mu_\psi \right) \;,
\end{equation}
leading to
\begin{align}
    \gamma_{\text{s}} & = \frac{\Gamma_{\text{sph}}}{f^2 T}\quad \sim (\alpha N_c)^5 \frac{T^3}{f^2} \;, \label{gammaS}\\
    \gamma_{\text{B}} & = \frac{12 \Gamma_{\text{sph}}}{T^3} \quad \sim (\alpha N_c)^5 T \;. \label{gammaR}
\end{align}

For the moment we assume that the displacement of $\varphi$ from its potential minimum and the resulting $\mu_\psi \neq 0$ are the only deviations from local thermal equilibrium. In particular, gauge interactions are so enough that the occupation numbers in the plasma can be described by Bose-Einstein and Fermi-Dirac distributions with slowly varying chemical potential and temperature at all times. To solve the system, \eqs \eqref{axionEOMFull} and \eqref{fermionEOMFull}
have to be complemented by suitable equations of motion for $H$ and $T$, for which we take
\begin{align}\label{RadiationBathEquation}
\gamma_{\text{s}} \dot{\tp}^2 & = \dot{\rhoR} + 4H \rhoR   %\;,\\
\end{align}
with \cite{Rubakov:2017xzr}
\begin{eqnarray}\label{EnergyDensities}
    \rhoR = g_\star \frac{\pi^2}{30} T^4 \ , \quad \rho_\varphi = \frac{\dot{\tp}^2}{2} + V \;,
\end{eqnarray}
and the standard Friedmann equation
\begin{eqnarray}
    3 M_P^2 H^2 & = \rho_\varphi + \rhoR \;.  \label{HubbleEOM}
\end{eqnarray}
Here $M_P$ is the reduced Planck mass, $g_\star$ corresponds to the number of 
degrees of freedom and in \eqs \eqref{RadiationBathEquation} and \eqref{EnergyDensities} we approximated $\braket{\dot{\varphi}^2} \approx \dot{\tp}^2$ as throughout.
In principle, there are corrections to \eqref{RadiationBathEquation} \cite{Laine:2021ego}, and the energy densities in \eqref{EnergyDensities} can be computed consistently within the Schwinger-Keldysh formalism \cite{Herranen:2013raa}. However, as argued in \cite{Laine:2021ego}, numerically the resulting corrections are small, and we neglect them for our purpose. 

For now, we shall moreover neglect Hubble dilution in \eqs \eqref{axionEOMFullHubble}  and \eqref{fermionEOMFullHubble} so that they become:
\begin{align}
-\ddot{\tp} & = V' - \frac{1}{f}  \left\langle q  \right\rangle   = V' +  \gamma_{\text{s}}  \dot{\tp} + \frac{\gamma_{\text{B}} T^2}{6 f} \mu_\psi\;,  \label{axionEOMFull}\\
\dot{\mu}_\psi &=  \frac{6}{T^2}\left\langle q \right\rangle  = - \frac{6f}{T^2}\left(\gamma_{\text{s}}  \dot{\tp} + \frac{\gamma_{\text{B}} T^2}{6 f} \mu_\psi\right)  \label{fermionEOMFull} \;.
\end{align}
Given the motivation from warm inflation or dark energy, our primary goal is to find stable (quasi)stationary solutions to the above equations in which $T$ is kept approximately constant (adiabatically tracking $\varphi$) over many $e$-folds of cosmic expansion by an equilibrium between Hubble expansion and sphaleron heating.\footnote{At the background level the widely-accepted common lore is that $T$ will either quickly approach zero due to the dilution of the plasma (returning to standard cold inflation or ending inflation) or grow until it reaches the stationary value. In principle feedback from perturbations and produced particles can spoil this simple picture, so that inflation may not be an attractor solution when fluctuations are taken into account.
In a well-known proposal of cold Abelian axion inflation \cite{Anber:2009ua}, such an instability has recently been discovered \cite{Peloso:2022ovc,vonEckardstein:2023gwk}. 
The robustness of warm inflation against feedback of this kind is to be verified in each specific model, as it requires knowledge of the particle content and their interactions. 
Since the focus of the present work lies on a general discussion of the difference that fermions can make in the limit of instantaneous thermalisation, we refrain from assessing in detail how realistic a sufficiently rapid thermalisation is in any specific model.
We remark, however, that the addition of fermions generically attenuates gauge fields and correspondingly their backreaction, which improves stability. Already in vacuum, this happens due to Schwinger pair production (see \cite{Domcke:2018eki,Domcke:2019qmm,Domcke:2021fee,Gorbar:2021rlt,Gorbar:2021zlr,Gorbar:2021ajq,Gorbar:2023zla} and \cite{Mirzagholi:2019jeb,Maleknejad:2019hdr,Maleknejad:2020yys,Maleknejad:2020pec} for the Abelian and non-Abelian cases, respectively), and the existence of real fermions in a thermal plasma can lead to a significantly stronger reduction of gauge fields. Thus, fermions may even attenuate a possible instability.}
The stationary solution of \eq \eqref{fermionEOMFull} is
\begin{equation} \label{qVanishing}
    \left\langle q  \right\rangle  = - f\left(\gamma_{\text{s}}   \dot{\tp} + \frac{\gamma_{\text{B}} T^2}{6 f} \mu_\psi\right) =0 \;,
\end{equation}
which implies in \eq \eqref{axionEOMFull} that $-\ddot{\tp}  = V'$, i.e., $\ddot{\tp}$ does not vanish asymptotically. As already pointed out in \cite{Berghaus:2020ekh}, there is no stationary solution in which sphaleron heating transfers energy from $\tp$ to the bath, as it would be required during warm inflation. The reason is that the sole source of chirality violation is the same process that drives the friction, and hence, whenever the \rhs of \eqref{fermionEOMFull} vanishes, the $\gamma_{\text{s}}  \dot{\tp}$-term in \eqref{axionEOMFull} is exactly compensated by the chemical potentials so that there is no dissipation. 

\subsection{Hierarchy of energy scales}\label{Sec:Scales}
We shall estimate the order of magnitude of some relevant energy scales, so that we can assess the importance of different effects in what follows.
First, we define  as customary \cite{Berera:2008ar}
\begin{equation} \label{Q}
Q \equiv \frac{\gamma_{\text{s}}}{3 H} \;.
\end{equation}
By definition, warm effects and sphaleron heating dominate over Hubble friction if $Q \gtrsim 1$ and we shall limit ourselves to this regime throughout.
In the stationary state, $\dot{\rhoR}=0$, the energy density \eqref{EnergyDensities} is determined by \eq \eqref{RadiationBathEquation}, 
which leads to 
\begin{equation}
 \dot{\tp}^2 =  \frac{2 \pi^2 g_\star}{45 Q} T^4\;, \label{radiationEOM}
\end{equation}

The $\varphi q/f$-interaction in \eqref{originalModel} represents a non-renormalisable dimension five operator and is usually thought of as an effective field theory (EFT) description that holds if
\begin{equation}
    \max\left(H,T\right) \lesssim f \;. \label{EFTConstraint}
\end{equation}
For instance, if the axion arises as a result of symmetry breaking, namely as phase of a PQ-field, then the radial mode of the PQ-field can only be neglected below the energy scale $f$. Thus, the validity of the effective field theory description is only guaranteed if \eqref{EFTConstraint} holds  \cite{Domcke:2019lxq}. 
 Comparing the rates \eqref{gammaS} and \eqref{gammaR}, we conclude that the condition $Q\gtrsim 1$ together with the requirement \eqref{EFTConstraint} imply the hierarchy
\begin{equation}\label{ScaleHierarchy}
    H \lesssim \gamma_{\text{s}} \lesssim \gamma_{\text{B}} \lesssim T \;.
\end{equation}
In the last inequality, we assumed that the collective coupling of the non-Abelian gauge theory is not strong, $\alpha N_c \lesssim 1$. 

Finally, we can estimate the order of magnitude of the chemical potential $\mu_\psi$. Assuming stationarity in \eqref{axionEOMFull} or \eqref{fermionEOMFull}, we get
\begin{equation} \label{magnitudeMuPsi}
    \frac{\mu_\psi}{T} \sim \frac{\dot{\tp}}{f T}\sim \sqrt{\frac{g_\star}{Q}}\frac{T}{f} \;,
\end{equation}
where we used \eq \eqref{radiationEOM} in the second step. Since $Q \gtrsim 1$ and $T \lesssim f$ (and for moderate $g_\star$), we generically have $\mu_\psi \lesssim T$.

\section{Concrete mechanisms for restoring sphaleron heating}\label{Sec:mechanisms}

The conclusion that sphaleron heating is disabled is correct if the dark sector\footnote{Here we use the term \emph{dark sector} for any set of fields that are all singlet under the SM gauge interactions. This sector may or may not contain Dark Matter candidates.} in which it operates comprises only the $SU(N_c)$ gauge fields, the axion $\varphi$, and massless fermions. 
In the following we argue that sphaleron heating can be re-activated if the dark sector remotely resembles the properties of the SM seen in Nature, or if it has any non-negigible interactions with the SM.

\subsection{Chirality changing interactions}
\label{ssec:mass}

As already noted in \cite{McLerran:1990de}, the simplest mechanism to avoid the cancellation \eqref{qVanishing} and to restore sphaleron heating is through chirality-violating interactions. In their presence, \eq \eqref{fermionEOMFull} becomes
\begin{align} \label{fermionEOMCh}
  \dot{\mu}_\psi &= -\frac{6 f}{T^2}\gamma_{\text{s}} \dot{\tp} - \gamma_{\text{B}} \mu_\psi - \gamma_{\text{ch}} \mu_\psi  \;,
\end{align}
and the equation of motion \eqref{axionEOMFull} yields
\begin{equation}
-\ddot{\tp} = V' +  \gamma_{\text{eff}} \dot{\tp}  \;,
\end{equation}
where the effective dissipation rate is
\begin{equation}\label{gamaEffChiralityviolation}
\gamma_{\text{eff}} = \frac{\gamma_{\text{ch}}}{\gamma_{\text{B}}+\gamma_{\text{ch}}} \gamma_{\text{s}}\;.
\end{equation}
For $\gamma_{\text{ch}} \rightarrow 0$, we recover $\gamma_{\text{eff}} \rightarrow 0$, in accordance with our previous consideration. In contrast, $\gamma_{\text{ch}}\gg \gamma_{\text{B}}$ leads to $\gamma_{\text{eff}} \approx \gamma_{\text{s}}$, the same result that one obtains in the absence of any fermions.\footnote{We remark that for sufficiently large $\gamma_{\text{ch}}$, the chiral current $n_\psi$ can no longer be regarded as approximately conserved and strictly speaking no chemical potential $\mu_\psi$ can be defined. However, this quantity may still be used to effectively parameterise the deviation from thermal equilibrium. Moreover,  
\eq \eqref{gamaEffChiralityviolation} yields the correct result in both limits $\gamma_{\text{ch}}\rightarrow \infty$
and
$\gamma_{\text{ch}}\rightarrow 0$. 
Hence, while a precision calculation would require more care,
the physical picture established here appears to be captured well by \eqref{gamaEffChiralityviolation}, irrespective of the precise interpretation of $\mu_\psi$.}

Probably the simplest way to introduce chirality flips is through a fermion mass term
\begin{equation}\label{VacuumMass}
    \mathcal{L} \supset m\bar{\psi}\psi \;.
\end{equation}
This leads to a rate \cite{Boyarsky:2020cyk}
\begin{equation}
    \gamma_{\text{ch}} = \frac{\kappa N_c \alpha   m^2}{T} \;, \label{gammaCh}
\end{equation}
where $\kappa$ is a coefficient of order $1$. Then sphaleron heating is restored if (see also \cite{Berghaus:2020ekh, Berghaus:2023ypi}) 
\begin{equation} \label{restorationMass}
\gamma_{\text{ch}} \gtrsim  \gamma_{\text{B}} \qquad \Longleftrightarrow \qquad m\gtrsim (\alpha N_c)^2 T \;,
\end{equation}
i.e., fermion masses need to be greater than the temperature, up to a possible suppression due to the gauge coupling constant. 
Note that it is sufficient if one  fermion species that couples to $G_\mu$ has a mass that fulfills \eqref{restorationMass}.

However, an explicit mass term is not even necessary, as
any chirality-violating interaction suffices to restore sphaleron heating \cite{McLerran:1990de}. As simple example, we can consider 
a Yukawa coupling\footnote{Note that a similar argument could me made with a coupling $y\bar{\psi}_L\psi_L^c$ or $y\bar{\psi}_R\psi_R^c$ if only a Weyl fermion $\psi_L$ or $\psi_R$ were present.}
\begin{equation} \label{Yukawa}
  \mathcal{L} \supset y \phi \bar{\psi}\psi \;,
\end{equation}
where $\phi$ is another scalar field and $y$ a coupling constant. 
Firstly, if $\phi$ develops a non-zero expectation value $\langle\phi\rangle$ during inflation, \eqref{Yukawa} can right away generate a (possibly time-dependent) mass term of the form \eqref{VacuumMass}.\footnote{\label{ThermalMassFootnote2}
Note that the thermal mass 
generated by the Yukawa interaction \eqref{Yukawa} 
is fundamentally different from \eqref{VacuumMass} in the sense that it modifies the dispersion relation, but does not flip chirality \cite{Kiessig:2010pr} (just as for gauge interactions, cf.~footnote \ref{ThermalMassFootnote}). In contrast to that, a non-zero $\braket{\varphi}$ generates a chirality violating mass term, just as the Higgs mechanism. 
}  
This may either occur because $\langle\phi\rangle$ has not relaxed to its minimum or because it has been displaced by quantum fluctuations.
Even if $\langle\phi\rangle=0$, the Yukawa interaction mediates chirality-violating scatterings.
The resulting rate of chirality violation scales as \cite{Bodeker:2019ajh}
\begin{equation} \label{rateYukawa}
     \gamma_{\text{ch}} \sim y^2 T \;.
\end{equation}
Thus even for a massless fermion $\psi$, sphaleron heating is restored if
\begin{equation} \label{restorationYukawa}
     \gamma_{\text{ch}} \gtrsim \gamma_{\text{B}} \qquad \Longleftrightarrow \qquad |y| \gtrsim (\alpha N_c)^{5/2} \;. 
\end{equation}

An interesting possibility for a chirality-violating interaction comes from a direct coupling to the axion:
\begin{equation} \label{YukawaAxion}
    \mathcal{L} \supset y_\varphi \varphi \bar{\psi}\psi \;.
\end{equation}
The background field value $\tp$ of the axion creates an effective fermion mass $m_{\text{eff}} = y_\varphi \tp$. Thus, it follows from \eq \eqref{restorationMass} that sphaleron heating is restored if
\begin{equation} \label{YukawaLowerBound}
    | y_\varphi \tp| \gtrsim (\alpha N_c)^2 T \;.
\end{equation}
Of course, the Yukawa interaction \eqref{YukawaAxion} is incompatible with an exact shift symmetry $\varphi \rightarrow \varphi + c$ of the axion. However, realistic models of warm inflation or dark energy  require in any case an axion potential $V(\varphi)$ that explicitly breaks the shift symmetry. In this case, the Yukawa interaction \eqref{YukawaAxion} can be regarded as viable as long as the mass resulting from it is smaller than the one generated by the potential. Since the coupling to fermions generates a thermal mass $\sim y_\varphi T$ for the axion \cite{Carrington:1991hz,Kiessig:2010pr},
we get the upper bound
\begin{equation} \label{YukawaUpperBound}
    |y_\varphi T| \lesssim  \sqrt{|V''\left(\tp\right)|} \;.
\end{equation}
Whether or not \eqs \eqref{YukawaLowerBound} and \eqref{YukawaUpperBound} can be fulfilled simultaneously depends on the concrete potential (see discussion in section \ref{sec:example} for a concrete example).

\subsection{Another sphaleron}
\label{ssec:sphaleron}

Another possibility for restoring sphaleron heating arises if the massless fermion $\psi$ is coupled to a second non-Abelian gauge field $\tilde{G}_\mu$, which does not directly interact with the axion, i.e., the model \eqref{originalModel} is extended as:
\begin{equation}
    \mathcal{L} = -\frac{1}{2} \text{Tr}\, G_{\mu\nu} G^{\mu\nu}  -\frac{1}{2} \text{Tr}\, \tilde{G}_{\mu\nu} \tilde{G}^{\mu\nu}  +\left(\frac{i}{2} \bar{\psi} \slashed{D} \psi + \text{h.c.}\right) -\frac{1}{2} \partial_\mu \varphi \partial^\mu \varphi + \frac{\varphi}{f} q - V(\varphi) \;.
\end{equation}
Here $\tilde{G}_{\mu\nu}$ is the field-strength tensor of $\tilde{G}_\mu$ and now the covariant derivative reads $D_\mu = \partial_\mu - i g G_\mu - i \tilde{g} \tilde{G}_\mu$, where $\tilde{g}$ is the new coupling constant. 

The gauge field $\tilde{G}_\mu$ comes with its own sphaleron effect. This leads to an additional contribution in the fermionic equation of motion so that \eq \eqref{fermionEOMFull} becomes
\begin{align} \label{fermionEOMSphaleron}
  \dot{\mu}_\psi &= -\frac{6 f}{T^2}\gamma_{\text{s}} \dot{\tp} - \gamma_{\text{B}} \mu_\psi - \tilde{\gamma}_{\text{B}} \mu_\psi   \;.
\end{align}
In analogy to \eq \eqref{gammaR}, we defined
\begin{equation}
   \tilde{\gamma}_{\text{B}} = \frac{12 \tilde{\Gamma}_{\text{sph}}}{T^3} \quad \sim (\tilde{\alpha} \tilde{N}_c)^5 T \;,
\end{equation}
where $\tilde{\alpha} = \tilde{g}^2/(4\pi)$ and $\tilde{N}_c$ is the number of colors in $\tilde{G}_\mu$. 
Comparing \eqs \eqref{fermionEOMSphaleron} and \eqref{fermionEOMCh}, we see that now $\gamma_{\text{ch}} = \tilde{\gamma}_B$. Therefore, sphaleron heating is restored if 
\begin{equation} \label{restorationSphaleron}
\tilde{\gamma}_{\text{B}} \gtrsim  \gamma_{\text{B}} \qquad \Longleftrightarrow \qquad \tilde{\alpha} \tilde{N}_c\gtrsim \alpha N_c \;.
\end{equation}
Note that it is sufficient that one fermion species that couples to $G_\mu$ is also charged under the second non-Abelian gauge group.
Given that the gauge group of the SM is a product group and most known fermions are charged under more than one interaction, this appears to be more natural than the absence of such couplings.

\subsection{Another source of chiral asymmetries}
\label{ssec:chiral}

A key assumption to reach the conclusion that sphaleron heating is disabled by massless fermions in section \ref{Sec:Disabeling} is that the axion represents the only source of chirality violation.
However, we emphasize that it is not necessary for $\psi$ itself to possess interactions that violate the total chirality, but it suffices that $\psi$ interacts through a chain of mediators with some fields that do (even if the mediator preserve the total chirality). Consider, for instance, a mediating interaction of the form $ (\bar{\psi_R}\psi_L)(\bar{\chi_L}\chi_R)$ with Weyl fermions $\chi_{L,R}$.  
This converts a $\psi$-particle with LH chirality and a $\chi$-particle with RH chirality into a 
$\chi$-particle with LH chirality and a $\psi$-particle with LH chirality.
hence conserving the total chirality, but violating $\bar{\psi}_L \gamma_0 \psi_L$ by two units. If $\chi_{L,R}$ have chirality-violating interactions, these will effectively be mediated to $\psi$.
There are multiple ways how chirality violation in other sectors could leak to the $\psi$, and one may argue that the presence of such interactions is more natural than their absence. 
There are also multiple ways of how chirality violation in other parts of the theory may occur. Again, since $C$ and $CP$ are both known to be violated in nature, the existence of such mechanisms in sectors that interact with $\psi$ appears to be more generic than their absence. 

For the sake of definiteness, we shall mention two specific examples. 
Firstly, the observable universe exhibits a matter-antimatter asymmetry \cite{Canetti:2012zc} at the level $\mu/T|_{\text{ext}} \sim 10^{-10}$ \cite{Planck:2018vyg}. 
Of course, one cannot expect that this asymmetry was preserved above the temperatures when the electroweak sphaleron was active \cite{Kuzmin:1985mm} (roughly 130 GeV \cite{DOnofrio:2014rug}) or when lepton number is violated (e.g.~in relation to the mechanism of neutrino mass generation, cf.~\cite{Asadi:2022njl,Agrawal:2021dbo} for a list of references), but it still serves as an estimate for the typical asymmetries that one may expect in Nature. 
In the simplest case, there is an efficient coupling of the system \eqref{originalModel} to
an external sector where a constant asymmetry $\mu/T|_{\text{ext}}$ is maintained by some other mechanism (in more complicated scenarios $\mu/T|_{\text{ext}}$  evolves dynamically).
If $\mu|_{\text{ext}}$ exceeds the value of $\mu_\psi$ induced by the sphaleron, then the fermionic chemical potential will generically by set by $\mu|_{\text{ext}}$ and the cancellation in \eq \eqref{qVanishing} is removed.
Comparison with \eq \eqref{magnitudeMuPsi} shows that a coupling to a non-vanishing $\mu/T|_{\text{ext}}$  would be sufficient to save sphaleron heating if
\begin{equation}\label{BAUrestoration}
    \sqrt{\frac{g_\star}{Q}}\frac{T}{f} \lesssim \frac{\mu}{T}\Big|_{\text{ext}} \;.
\end{equation}
In principle, this can be achieved with large $Q$ or small $T/f$. 

Another intriguing possibility 
which requires no additional deviations from equilibrium 
is that the axion itself produces the required asymmetries. 
A concrete realization of this idea can be implemented in the setup of \cite{Anber:2015yca,Adshead:2016iae,Jimenez:2017cdr,Domcke:2019mnd,Domcke:2020quw,Domcke:2022kfs}, when the axion additionally couples to the hypercharge gauge field of the SM,\footnote
{Even without adding an Abelian gauge field, it is conceivable that lepton asymmetries are generated during inflation by the gravitational anomaly \cite{Maleknejad:2016dci}.}
\begin{equation} \label{hyperchargeCoupling}
    \mathcal{L} \supset \frac{g_Y^2}{16 \pi^2} \frac{\varphi}{f} Y_{\mu\nu}\tilde{Y}^{\mu\nu} \;,
\end{equation}
where $g_Y$ is the hypercharge coupling and $Y_{\mu\nu}$ and $\tilde{Y}_{\mu\nu}$ are the corresponding field strength and its dual, respectively. In \cite{Domcke:2022kfs}, cold Abelian axion inflation was considered\footnote{Also in the absence of thermal effects, an axion can drive inflation \cite{Freese:1990rb}, and this possibility has been discussed both for Abelian \cite{Anber:2009ua} and non-Abelian \cite{Maleknejad:2011jw,Maleknejad:2011sq,Adshead:2012kp} gauge groups; see \cite{Maleknejad:2012fw} for a review. We remark that thermalization is harder to achieve in the Abelian case due to the absence of sphaleron effects \cite{Domcke:2019qmm,Mirbabayi:2022cbt}.} and in this case, the coupling \eqref{hyperchargeCoupling} leads to the  dual production of hypermagnetic helicity and fermionic charge asymmetries
on the order of
\begin{equation}
    \frac{\mu_i}{T} = - 6 \epsilon_i Y_i^2 \upsilon \;,
\end{equation}
where $Y_i$ is the hypercharge and $\epsilon_i$ distinguishes between left- and right-handed fermions. Moreover, $\upsilon$ 
is a dimensionless yield parameter. Computing it is technically challenging since it requires solving the nonlinearly coupled system of axion, hypercharge gauge field and possibly fermions \cite{Anber:2009ua,Barnaby:2011vw,Cook:2011hg,Barnaby:2011qe,Pajer:2013fsa,Notari:2016npn,Domcke:2018eki,Lozanov:2018kpk,DallAgata:2019yrr,Sobol:2019xls,Domcke:2019qmm,Domcke:2020zez,Domcke:2021fee,Gorbar:2021rlt,Gorbar:2021zlr,Gorbar:2021ajq,Caravano:2021bfn,Caravano:2022epk,Domcke:2022kfs,Peloso:2022ovc,Garcia-Bellido:2023ser,Gorbar:2023zla,Figueroa:2023oxc,Domcke:2023tnn,vonEckardstein:2023gwk}.\footnote{Interestingly, fermions also play a crucial role in cold Abelian axion inflation \cite{Domcke:2018eki,Domcke:2019qmm,Domcke:2021fee,Gorbar:2021rlt,Gorbar:2021zlr,Gorbar:2021ajq,Gorbar:2023zla}: Fermion production due to the Schwinger effect reduces the gauge fields. In turn, this implies that the backreaction of the gauge field on the axion is attenuated.}
Values $\upsilon \sim 10^{-7 \ldots -8}$ were considered in \cite{Domcke:2022kfs}. Now sphaleron heating is restored if two conditions are fulfilled. First, some of the SM fermions need to be coupled efficiently to the spectator $\psi$ (or alternatively to the hypermagnetic field).
Second, the chemical potentials in the SM $\mu_i$ must be sizable as compared to $\mu_\psi$, i.e., (see \eq \eqref{magnitudeMuPsi})
\begin{equation}  \label{chiBound}
    \upsilon \gtrsim \sqrt{\frac{g_\star}{Q}}\frac{T}{f} \;.
\end{equation}
In this case, the axion itself can save sphaleron heating. Of course, the numerical values derived in \cite{Domcke:2022kfs} will need to be revised quantitatively because they were obtained under the assumption that the universe was cold and empty during inflation, while the thermal plasma in warm inflation scenarios can lead to a larger washout of the helical hypermagnetic fields \cite{Figueroa:2019jsi,Boyarsky:2020cyk}.
Such computations are technically challenging, and even in the cold inflation case the effect of backreaction is not fully understood \cite{Anber:2009ua,Barnaby:2011vw,Cook:2011hg,Barnaby:2011qe,Pajer:2013fsa,Notari:2016npn,Domcke:2018eki,Lozanov:2018kpk,DallAgata:2019yrr,Sobol:2019xls,Domcke:2019qmm,Domcke:2020zez,Domcke:2021fee,Gorbar:2021rlt,Gorbar:2021zlr,Gorbar:2021ajq,Caravano:2021bfn,Caravano:2022epk,Domcke:2022kfs,Peloso:2022ovc,Garcia-Bellido:2023ser,Gorbar:2023zla,Figueroa:2023oxc,Domcke:2023tnn,vonEckardstein:2023gwk}.
However, qualitatively this offers a mechanism how the displacement of $\langle\varphi\rangle$ from its ground state itself can source the generation of chiral asymmetries.

\subsection{Hubble dilution}
\label{ssec:dilution}

So far, we have neglected Hubble dilution in \eqref{axionEOMFull} and \eqref{fermionEOMFull}. Now we turn back to the full equations of motion \eqref{axionEOMFullHubble} and \eqref{fermionEOMFullHubble}. As compared to the approximations \eqs \eqref{axionEOMFull} and \eqref{fermionEOMFull}, we see that the correction due to $H$ in \eqref{axionEOMFullHubble}  is suppressed by
\begin{equation} \label{suppressionQ}
   \frac{3 H}{\gamma_{\text{s}}} = \frac{1}{Q} \;.
\end{equation}
By construction, the effect of Hubble friction in \eqref{axionEOMFullHubble} is sub-dominant in the warm inflation regime $Q\gtrsim 1$. In \eq \eqref{fermionEOMFullHubble}, Hubble dilution leads to the stationary solution
\begin{equation}\label{MuPsiWithH}
    \mu_\psi = - \frac{6 f}{T^2} \frac{\gamma_{\text{s}} \dot{\tp}}{\gamma_{\text{B}}+3H} \;,
\end{equation}
which gives in \eq \eqref{axionEOMFullHubble}: 
\begin{equation}\label{AxionEoMWithHstationary}
    -\ddot{\tp}  = V' + 3 H\left(1 + \gamma_{\text{s}}  \frac{1}{3H+\gamma_{\text{B}}}\right)\dot{\tp} \;.
\end{equation}
Thus, the effective damping in this case is
\begin{equation}\label{gammaEffHubble}
\gamma_{\text{eff}} = \frac{3H}{\gamma_{\text{B}}+3H} \gamma_{\text{s}} \simeq \frac{T^2}{12 f^2 Q} \gamma_{\text{s}}.
\end{equation}
Comparing to \eq \eqref{gamaEffChiralityviolation}, we see that Hubble expansion plays the same role as chirality violating interactions. 
In contrast to the complete cancellation observed in section \ref{Sec:Disabeling}, this rate now is non-zero.
However, as compared to the case without fermions, 
it is suppressed by
\begin{equation} \label{suppressionHubble}
    \frac{3H}{\gamma_{\text{B}}} = \frac{\gamma_{\text{s}}}{Q \gamma_{\text{B}}} \sim \frac{1}{Q} \frac{T^2}{f^2} 
    < \frac{1}{Q} < 1
\end{equation}
where we used \eqs \eqref{gammaS} and \eqref{gammaR}  for the estimate.
Comparing with \eqref{suppressionQ}, 
we see that there is an additional suppression of $T^2/f^2$. 
Thus, Hubble dilution plays no role in \eq \eqref{fermionEOMFullHubble}, which  justifies that we neglected this effect in the previous discussions.

\subsection{Cumulative effects}

We have seen that the suppression of the sphaleron heating mechanism can be alleviated if, in addition to the axion coupling, an additional source of chirality violation is present either in the sector where the sphaleron acts or in a sector that is connected to it by mediator interactions.\footnote{The role of mediators that themselves preserve the total chirality by connecting $\psi$ to sectors where it is violated may be compared to the role of spectator processes in baryogenesis \cite{Buchmuller:2001sr}.}
It is important to note that this effect is additive, i.e., it can also arise as cumulative result of many small chirality violations distributed over many fields. 
We illustrate this for the minimal and unavoidable mechanism, namely the dilution of chemical potentials by Hubble expansion.

Extending \eqs \eqref{axionEOMFullHubble} and \eqref{fermionEOMFullHubble}, we add fields $\chi_i$, $i=1,\ldots,N_f$, which are coupled to $\psi$ (but not to $\varphi$). This leads to equations of motion of the form
\begin{align}
-\ddot{\tp} & = V' + \left(3H+\gamma_{\text{s}} \right) \dot{\tp} +   \frac{\gamma_{\text{B}} T^2}{6 f}  \mu_\psi \;, \label{RepeatedAxionEoM} \\
\dot{\mu}_\psi &= - \frac{6 f}{T^2}\left(\gamma_{\text{s}} \dot{\tp} +  \frac{\gamma_{\text{B}} T^2}{6 f} \mu_\psi \right) - 3 H \mu_\psi  - \sum_{i=1}^{N_f}\tilde{\gamma}_i \left(\mu_\psi-\mu_{\chi_i}\right) \;,\label{Mupsiwithspectators}\\
\dot{\mu}_{\chi_i} & =  \tilde{\gamma}_i \left(\mu_\psi-\mu_{\chi_i}\right) - 3 H \mu_{\chi_i} \;,\label{SpectatorEquilibration}
\end{align}
where crucially we included Hubble dilution in the latter two equations. 
The details of the interactions that connect the $\chi_i$ to each other and to $\psi$ are not relevant for main point here, all that matters is that they drive all chemical potentials to the same value $\mu_\psi = \mu_{\chi_1} = \ldots = \mu_{\chi_{N_f}}$ with some rates $\tilde{\gamma}_i > H$.\footnote{
In principle, the network of kinetic equations is more complicated, as the rates $\tilde{\gamma}_{ij}$ at which different chemical potentials $\mu_{\chi_i}$ and $\mu_{\chi_j}$ equilibrate can be different from each other and can differ from the rate $\tilde{\gamma}_{i\psi}$ at which they equilibrate with $\mu_\psi$. Here we for simplicity assume that all $\tilde{\gamma}_{i\psi}, \tilde{\gamma}_{ij} > H$, so that all 
$\mu_{\chi_i}$ track the value of $\mu_\psi$.
} 
Stationarity in \eq \eqref{SpectatorEquilibration} implies
\begin{equation} \label{muHubbleChi}
    \mu_{\chi_i} = \frac{\mu_\psi}{1 + \frac{3H}{\tilde{\gamma}_i}} \;.
\end{equation}
Insertion into \eqref{Mupsiwithspectators} 
permits to find the stationary value for $\mu_\psi$ in terms of other parameters. For simplicity, we assume that all the $\tilde{\gamma}_i$ are of the same order and expand in $H/\tilde{\gamma}_i$ to obtain
\begin{equation}  \label{muHubblePsi}
    \mu_\psi = -\frac{6 f}{T^2} \frac{\gamma_{\text{s}} \dot{\tp}}{\gamma_{\text{B}}+3H(1+N_f)} \;.
\end{equation}
In \eq \eqref{RepeatedAxionEoM}, this implies 
\begin{eqnarray}\label{specatorsSaveSphaleronHeating}
      -\ddot{\tp}  = V' + 3H \dot{\tp} + \gamma_{\text{s}}\dot{\tp}\left(1 - \frac{\gamma_{\text{B}}}{3H(1+N_f)+\gamma_{\text{B}}}\right) \ ,
\end{eqnarray}
leading to 
\begin{equation}\label{gammaEffHubbleNf}
\gamma_{\text{eff}} = \frac{3H(1+N_f)}{\gamma_{\text{B}}+3H(1+N_f)} \gamma_{\text{s}} \;.
\end{equation}
Hence, the suppression of sphaleron heating is overcome if $3H(1+N_f) \gtrsim \gamma_{\text{B}}$ 
or\footnote{An unavoidable upper bound on the number of hidden species is $10^{32}$ \cite{Dvali:2007hz,Dvali:2007wp,Dvali:2008ec}.}
\begin{equation} \label{speciesHubble}
    N_f \gtrsim Q \frac{f^2}{T^2} \;.
\end{equation}
We see that even the unavoidable existence of Hubble dilution can reactive sphaleron heating.

It is important to note that the $\chi_i$ do not have to couple directly to $\psi$ -- the existence of an effective channel for particle number exchange between $\chi_i$ and $\psi$ suffices. 
Moreover, it is very interesting that the effect of Hubble dilution becomes important for $f/T$ close to $1$. Therefore, it can be complimentary to mechanisms such as the one shown in \eq \eqref{chiBound}, which are suppressed at large temperatures.
Finally,  similar cumulative effects can be observed if several of the other mechanisms that can reactivate sphaleron heating coexist.\footnote{As the simplest example, one could consider a system of $N_f$ coupled fermions where each field has a small mass $m\sim (\alpha N_c)^2 T/N_f$, \cf \eq \eqref{restorationMass}.}

\section{Example: warm inflation with monomial potential}
\label{sec:example}

Next we shall discuss the concrete example of a monomial potential,
\begin{equation} \label{examplaryPotential}
    V = C^{4-\beta} \varphi^\beta \;,
\end{equation}
as also studied in \cite{Berghaus:2020ekh,Laine:2021ego}. $C$ has dimension of energy and for definiteness, we shall restrict ourselves to $1\leq \beta \leq 3$.
In order to guarantee validity of effective field theory and ensure a definite hierarchy of energy scales, we will impose $f<M_P$ and $C<M_P$.
Throughout, we shall assume that the system is already thermalized (see \cite{Ferreira:2017lnd,Ferreira:2017wlx,Berghaus:2019whh,Laine:2021ego,DeRocco:2021rzv,Laine:2022ytc} for partially controversial discussions about how a thermal state can be reached).

\subsection{Energy scales without fermion}

In a first step, we will neglect the contribution of fermions. Combining \eqs \eqref{gammaS}, \eqref{HubbleEOM}, \eqref{axionEOMFull}, \eqref{Q}, and \eqref{radiationEOM},  we can express temperature as a function of the field value $\tp$ and constants of the theory (see also \cite{Berghaus:2020ekh}),
\begin{equation} \label{temperature}
 T=\left(  \frac{15 \sqrt{3} \beta ^2 f^2 \tp ^{\frac{3 \beta }{2}-2} C^{6-\frac{3
   \beta }{2}} M_P}{2 \pi ^2 c_{\text{sph}}  g_\star (\alpha N_c)^5  }\right)^{1/7} \;,
\end{equation}
where $c_{\text{sph}}$ is defined in \eq \eqref{GammaSph} and we moreover used \eq \eqref{EnergyDensities} in the approximation $\rho_\varphi\approx V \gg \rhoR$. Already here we see an important difference depending on whether $\beta$ is smaller or larger than $4/3$. For $\beta<4/3$, the temperature increases with decreasing field value $\tp$. In contrast, $T$ gets lower as $\tp$ goes down in the case $\beta > 4/3$.

Next we evaluate $Q$ and the slow-roll parameter 
\begin{eqnarray}\label{SlowRollEpsilon}
    \epsilon \equiv \frac{M_P^2}{2(1+Q)}\left(\frac{V'}{V}\right)^2 
    \simeq 
\frac{M_P^2}{2Q}\left(\frac{V'}{V}\right)^2  \;,
\end{eqnarray}
which gives
\begin{align}
Q&= \left(\frac{375 c_{\text{sph}}^4 (\alpha N_c)^{20} \beta ^6    C^{4-\beta
   } M_P^{10}}{8 \pi ^6 g_\star^3 \tp ^{6-\beta} f^8 }\right)^{1/7}\;, \label{QEx} \\
 \epsilon &= \left(\frac{\pi ^6 \beta ^8 g_\star^3 f^8    M_P^4}{6000 c_{\text{sph}}^4
   (\alpha N_c)^{20} \tp^{\beta +8} C^{4-\beta}}\right)^{1/7}\;. \label{epsilonEx} 
\end{align}
We see that both $Q$ and $\epsilon$ grow as $\tp$ decreases. Now parameter have to be chosen in such a way that two requirements are fulfilled.

First, we want to focus on a regime where thermal friction dominates. 
Since $Q$ grows with time, the minimal condition comes from the point where $\epsilon$ crosses $1$ and so the system exits the quasi-de Sitter state. The corresponding field value is
\begin{equation} \label{phiEnd}
    \tp_{\text{end}} = f \left(\frac{\pi ^6 \beta ^8 g_\star^3  M_P^4}{6000 (\alpha
   N_c)^{20} c_{\text{sph}}^4 C^{4-\beta} f^\beta}\right)^{1/(8+\beta)} \;.
\end{equation}
Demanding that $Q>1$ for $\tp = \tp_{\text{end}}$ implies 
\begin{equation} \label{Cmin}
      C > C_{\text{min}} = f \left(\frac{2^{\beta/2}  \pi^{6} g_\star^3 }{3\cdot 5^3  \beta ^{ \beta } c_{\text{sph}}^4 (\alpha N_c)^{20} }\right)^{1/(4-\beta)}  \left(\frac{f}{M_P}\right)^{(4+\beta)/(4-\beta)} \;. % \sim f \left(\frac{f}{M_P}\right)^{(4+\beta)/(4-\beta)} 
\end{equation}
For a given $C$, the largest field value for which still $Q>1$ is
\begin{equation} \label{fieldMaxQ}
    \tp_{\text{max,Q}} = f \left(\frac{375 c_{\text{sph}}^4 (\alpha N_c) ^{20} \beta ^6  C^{4-\beta } M_P^{10}}{8
   \pi ^6 f^{14-\beta} g_\star^3}\right)^{1/(6-\beta)} \;.
\end{equation}
As it should, $\tp_{\text{max,Q}}$ coincides with $\tp_{\text{end}}$ for $C=C_{\text{min}}$. 

The next condition on parameter choices comes from validity of EFT (c.f. \eq \eqref{EFTConstraint}). Imposing $T/f<1$ at $\tp=\tp_{\text{end}}$
implies
\begin{equation} \label{Cmax}
    C  < C_{\text{max}} = f \left(\frac{2^{\beta/2 } \pi ^{2-\beta/2}
 c_{\text{sph}}^{\beta/2 } g_\star^{(4-\beta)/
4} (\alpha  N_c)^{5 \beta/2 }}{5^{(4-\beta)/4} 3 \beta ^{\beta }}      \right)^{1/(4-\beta )} \left(\frac{f}{M_P}\right)^{\beta/(2 (4-\beta ))} \;.
\end{equation} 
Evidently, a viable parameter choice has to fulfill $C_{\text{min}}<C_{\text{max}}$. Since $\beta/2<\beta +4$, comparison with \eq \eqref{Cmin} shows that we can always achieve this by choosing $f/M_P$ sufficiently small so that we get the consistent hierarchy of energy scales
\begin{equation} \label{hierarchies}
    C_{\text{min}}<C< C_{\text{max}}<f<M_P \;.
\end{equation}
We will later show realistic parameter values that implement \eq \eqref{hierarchies}.

Importantly, \eq \eqref{hierarchies} represents a conservative bound in the case $\beta<4/3$. As the temperature \eqref{temperature} decreases with the field value $\tp$, the EFT description is valid throughout the exponential expansion if it is valid at its end where $\tp=\tp_{\text{end}}$.
So the only upper bound on $\tp$ comes from the point where thermal effects no longer dominate (c.f. \eq \eqref{fieldMaxQ}), i.e.,
\begin{equation}
    \beta<4/3: \qquad \tp_{\text{end}} < \tp < \tp_{\text{max,Q}} \;.
\end{equation}
In contrast, temperature grows with $\tp$ for $\beta > 4/3$. Therefore, the validity of the EFT description is only guaranteed for $\tp <  \tp_{\text{max,T}}$, where
\begin{equation} \label{fieldMaxT}
   \tp_{\text{max,T}} = f \left(\frac{15 \sqrt{3} \beta ^2 C^{6-\frac{3 \beta }{2}} M_P}{2 \pi ^2 c_{\text{sph}} (\alpha N_c)^5  g_\star f^{7-\frac{3 \beta }{2}} }\right)^{2/(4-3\beta)} \;.
\end{equation}
As it should, $\tp_{\text{max,T}}=\tp_{\text{end}}$ 
for $C=C_{\text{max}}$.
In summary, the viable field range for $\tp$ is
\begin{equation}
    \beta > 4/3: \qquad \tp_{\text{end}} < \tp < \min\left(\tp_{\text{max,Q}}, \tp_{\text{max,T}}\right) \;.
\end{equation}

\begin{figure}
    \centering
    \begin{subfigure}{0.47\textwidth} 
	\centering
	\includegraphics[width=\textwidth]{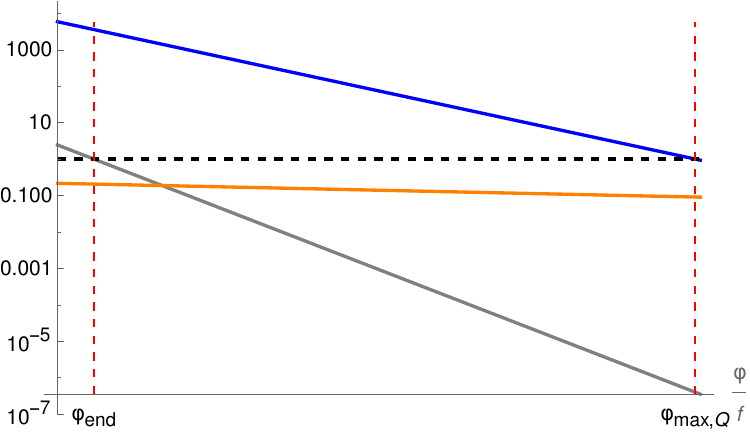}
	\caption{The case $\beta=1$, when $T/f$ decreases as a function of $\tp$, for the parameter choice \eqref{parametersBeta1}.}
 \label{fig:dynamicsBeta1}
\end{subfigure}
\hspace{0.04\textwidth}
\begin{subfigure}{0.47\textwidth}
	\centering
	\includegraphics[width=\textwidth]{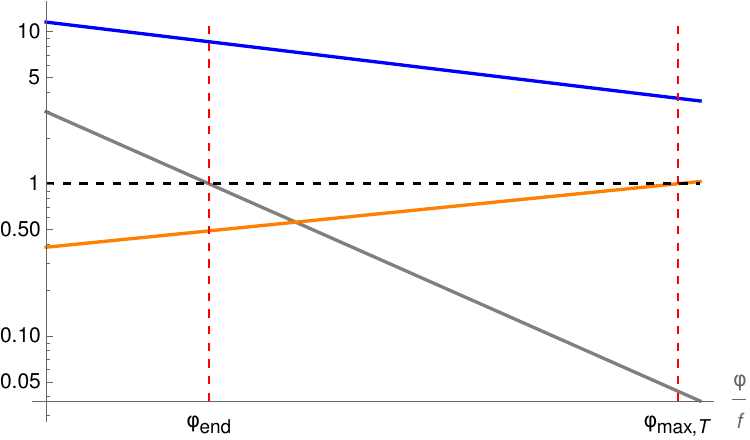}
	\caption{The case $\beta = 3$, when $T/f$ increases as a function of $\tp$, for the parameter choice \eqref{parametersBeta3}.}
  \label{fig:dynamicsBeta3}
\end{subfigure}
    \caption{Plot of dynamics resulting from the potential \eqref{examplaryPotential}, where the field value $\tp$ decreases as time progresses. The slow-roll parameter $\epsilon$ is shown in grey, and $Q$ and $T/f$ are displayed in blue and orange, respectively. The point where $\epsilon$ crosses $1$ marks $\tp =\tp_{\text{end}}$ when the systems exits a quasi-de Sitter stage, while above $\tp =\tp_{\text{max,Q}}$ the quantity $Q$ falls below $1$ and above $\tp =\tp_{\text{max,T}}$ the temperature $T$ exceeds $f$. In other words, the region $\tp<\tp_{\text{end}}$ does not lead to exponential expansion, whereas for $\tp>\tp_{\text{max,Q}}$ thermal dissipation is not strong. We remark that for $\beta > 4/3$, the field value $\tp_{\text{max,Q}}$, where $Q$ crosses $1$, could also be lower than $\tp_{\text{max,T}}$.}
    \label{fig:dynamics}
\end{figure}

Additionally, we can compare the temperature $T$ and the scale $C$ of the potential. Evaluating this quantity at $\tp = \tp_{\text{end}}$, we get
\begin{equation} \label{TOverC}
    \frac{T}{C}\Big\vert_{\tp = \tp_{\text{end}}} =\left(\frac{5^{(4-\beta)/2} 9  \beta ^{2 \beta } }{2^{\beta } \pi ^{4-\beta} c_{\text{sph}}^{\beta }   g_\star^{(4-\beta)/2}(\alpha  N_c)^{5
   \beta }}    
  \right)^{1/(8+\beta)} \left(\frac{f^{2\beta} M_P^\beta}{C^{3\beta}} \right)^{1/(8+\beta)} \;.
\end{equation}
Since $C<f<M_P$ and $\alpha N_c <1$, we see that $T>C$ for generic parameter choices.
This resonates with the basic idea of the axion as (pseudo-)Nambu Goldstone boson, where the symmetry breaking due to the potential should be small. 

Finally, we shall consider two examples with concrete parameter values. The first one is defined by
\begin{equation} \label{parametersBeta1}
    \beta = 1 \;, \quad f=10^{-6} M_P \;, \quad C=10^{-8} M_P \;, \quad g_\star = 10  \;, \quad \alpha = 0.2  \;, \quad N_c = 3 \;.
\end{equation}
For the following estimates, we shall moreover set $c_{\text{sph}}\approx 1$, although an accurate analysis reveals that $c_{\text{sph}}$ also contains a logarithmic dependence on model parameters and dynamical quantities; see \cite{Laine:2021ego} for details.
Evaluating \eqs \eqref{Cmin} and \eqref{Cmax}, we see that $C_{\text{min}}\approx 5\cdot 10^{-14} M_P < C < C_{\text{max}}\approx 1\cdot 10^{-7} M_P$, in accordance with the hierarchy \eqref{hierarchies}. From \eqs \eqref{phiEnd} and \eqref{fieldMaxQ} we get $\tp_{\text{end}}\approx 10^{-2} M_P$ and $\tp_{\text{max,Q}}\approx 10^3 M_P$. Finally, \eq \eqref{TOverC} shows that $T/C\vert_{\tp = \tp_{\text{end}}} \approx 21 > 1$. The dependence of the $\epsilon$, $Q$ and $T$ on the field value $\tp$ is shown in \fig \ref{fig:dynamicsBeta1}.

As second exemplary parameter choice is
\begin{equation} \label{parametersBeta3}
    \beta = 3 \;, \quad f=10^{-3} M_P \;, \quad C=10^{-12} M_P \;, \quad g_\star = 10  \;, \quad \alpha = 0.2  \;, \quad N_c = 3 \;.
\end{equation}
Now we have $C_{\text{min}}\approx 7\cdot 10^{-18} M_P < C < C_{\text{max}}\approx  5\cdot 10^{-11} M_P$ as well as $\tp_{\text{end}}\approx 7\cdot 10^{-1} M_P$, $\tp_{\text{max,Q}}\approx 10^2 M_P$, and $\tp_{\text{max,T}}\approx 5 M_P$, where we also took into account \eq \eqref{fieldMaxT}, and additionally $T/C\vert_{\tp = \tp_{\text{end}}} \approx 5 \cdot 10^8$. For the values \eqref{parametersBeta3}, the dependence of the relevant quantities on $\tp$ is depicted in \fig \ref{fig:dynamicsBeta3}.

\subsection{Order of magnitude of fermionic effects}
Now that we know how different energy scales are related to each other, we determine which of the mechanisms for the restoration of sphaleron heating is relevant in a realistic scenario of exponential expansion due to an axion. Throughout this subsection, we  leave out all dimensionless parameters and thereby only focus on ratios of energies. Then we get from \eqs \eqref{Cmin} and \eqref{Cmax}
\begin{equation}
C_{\text{min}} \sim f  \left(\frac{f}{M_P}\right)^{(4+\beta)/(4-\beta)} 
\;,
\qquad
    C_{\text{max}} \sim f \left(\frac{f}{M_P}\right)^{\beta/(2 (4-\beta ))} \;,
\end{equation}
where the symbol $\sim$ indicates that we ignore all numerical factors and constants ($g_\star, \alpha, N_c$, etc.).
In this way, we can determine which mechanisms are naturally operative as a result of energy hierarchies that are inevitably present. Evidently, we cannot exclude that our findings change in a concrete scenario if some of the dimensionless parameters are chosen to be sufficiently small (or large). 

The efficiency of both a chiral mass and an external chemical potential depends on temperature (c.f. \eqs~\eqref{restorationMass} and \eqref{BAUrestoration}). For $\beta<4/3$, it is in the range
\begin{equation}
    \left(\frac{C^{\beta }  f^{3 \beta
   -4}}{C_\text{max}^{4 (\beta-1)}}\right)^{(4-\beta)/((6-\beta ) \beta)}\lesssim \frac{T}{f} \lesssim \left(\frac{C}{C_{\text{max}}}\right)^{(8-2\beta)/(8+\beta)} \;,
\end{equation}
where the lower bound comes from the requiring $Q>1$, i.e., $\tp = \tp_{\text{max,Q}}$, while the upper bound is due to $\tp = \tp_{\text{end}}$. For the example \eqref{parametersBeta1}, we have $T(\tp_{\text{max,Q}})\approx 0.09f$ and $T(\tp_{\text{end}})\approx 0.2f$.
In the case $\beta > 4/3$, we get
\begin{equation}
\left(\frac{C}{C_{\text{max}}} \right)^{(8-2\beta)/(8+\beta)} \lesssim \frac{T}{f} \lesssim   \min\left(1, \left(\frac{C^{\beta }  f^{3 \beta
   -4}}{C_\text{max}^{4 (\beta-1)}}\right)^{(4-\beta)/((6-\beta ) \beta)} \right)\;,
\end{equation}
where now the upper bound comes from the minimum of  $\tp = \tp_{\text{max,Q}}$ and $\tp = \tp_{\text{max,T}}$ while the lower bound is due to $\tp = \tp_{\text{end}}$. The example \eqref{parametersBeta3} yields $T(\tp_{\text{end}})\approx 0.5f$ while $T(\tp_{\text{max,T}})=f$ by construction. Since a chiral mass and an external chemical potential are more effective at low temperatures, we see that their relevance decreases with time in the case $\beta<4/3$ while it increases $\beta > 4/3$.

The efficiency of Hubble dilution (c.f. \eq \eqref{speciesHubble}) is set by
\begin{equation} \label{speciesHubbleEv}
Q\frac{f^2}{T^2} \sim \left(\frac{f^2 M_p^8}{\phi ^{2 (\beta +1)} C^{2 (4-\beta )}}\right)^{1/7} \;.
\end{equation}
Therefore, this effect is always more relevant at initial times than towards the end of the quasi-de Sitter stage. In the example \eqref{parametersBeta1}, $\approx 3\cdot 10^3$ fields are efficient at $\tp=\tp_{\text{max,Q}}$ while $\approx 2\cdot 10^6$ would be required at $\tp=\tp_{\text{end}}$. For the choice \eqref{parametersBeta3}, the corresponding numbers are $\approx 55$ at $\tp=\tp_{\text{max,T}}$ and $\approx 5.4\cdot 10^2$ at $\tp=\tp_{\text{end}}$.

Finally, we turn to the direct coupling \eqref{YukawaAxion} of the axion to the fermion. As discussed, this coupling must not be too big in order to ensure that the resulting breaking of the axionic shift symmetry is not stronger than the corresponding effect due to the potential. It turns out that this condition, which is formulated in \eq \eqref{YukawaUpperBound}, leads to a crucial constraint. Evaluating\footnote
{Instead of \eq \eqref{YukawaUpperBound}, one could impose the criterion
	\begin{equation*} 
		|y_\varphi \tp| \lesssim  \frac{\sqrt{V\left(\tp\right)}}{T} \;,
	\end{equation*}
in which the energy densities resulting from the thermal mass and the vacuum potential are compared. Interestingly,  $\tp_{\text{end}} T/\sqrt{V(\tp_{\text{end}})}$ leads to the same scaling as shown in \eq \eqref{YukawaUpperBoundEv}.}
\begin{equation} \label{YukawaUpperBoundEv}
 \frac{T}{\sqrt{|V''(\tp_{\text{end}})}|} \sim   \left(\frac{f^{2 } M_p}{C^{3 } }\right)^{(4-\beta)(\beta +8)}\;,
\end{equation}
we see that keeping  $T \ll \sqrt{|V''(\tp_{\text{end}})|}$ requires $C>f$ whereas the bound \eqref{Cmax} implies $C<C_{\text{max}}<f$. Thus, it seems difficult to keep the breaking of the axionic shift symmetry small while guaranteeing the validity of the EFT-description. Indeed, we confirm that $T \gg \sqrt{|V''(\tp_{\text{end}})|}$ in both examples \eqref{parametersBeta1} and \eqref{parametersBeta3}.

Summarizing, we can categorize the different mechanisms for restoring sphaleron heating two categories:
\begin{itemize}
    \item \emph{Static mechanisms.}
    Chirality violating interactions due to a direct Yukawa coupling to another field (see \eq \eqref{restorationYukawa}) or a second sphaleron (see \eq \eqref{restorationSphaleron}) are independent of $T$ and $\tp$. Instead, their efficiency is solely determined by the comparison of the values of fundamental constants, comparing the newly added coupling constant to the interaction strength $\sim \alpha N_c$ of the sphaleron driving the exponential expansion.
    \item \emph{Dynamical mechanisms.}
    The chiral mass (see \eq \eqref{restorationMass}) and the external chemical potential (see \eq \eqref{BAUrestoration}) work better with lower temperatures. Therefore, their importance decreases over time for $\beta<4/3$ while it increases in the opposite case. For all $1\leq \beta \leq 3$, the efficiency of Hubble dilution (see \eq \eqref{speciesHubble}) diminishes in the course of evolution. Therefore, the aforementioned effects are complimentary for $\beta >4/3$, with chiral mass/external chirality violation and Hubble dilution becoming important for late and early times, respectively. A special case of a dynamical mechanism is a direct Yukawa coupling of fermions to the axion. For the monomial potential \eqref{examplaryPotential}, it seems to be disfavored, as long as the breaking of the axionic shift symmetry is kept small (c.f. \eqs \eqref{YukawaAxion} and \eqref{YukawaUpperBound}).   
\end{itemize}

\subsection{Outlook: new mechanisms for ending warm inflation}
The efficiency of some of the mechanisms discussed in section \ref{Sec:mechanisms} only depends on fundamental coupling constants. Examples include scatterings mediated by Yukawa interactions and the presence of a second sphaleron, see \eqref{restorationYukawa} and \eqref{restorationSphaleron}.
In other cases, the effective dissipation coefficient $\gamma_{\text{eff}}$ depends on $T$ and $\varphi$. For instance,
the chiral mass and the external chemical potential work better with lower temperatures, cf.~\eqref{restorationMass} and \eqref{BAUrestoration}.
Likewise, the right hand side of \eqref{speciesHubble} is a function of $T$ and $\varphi$. 
This implies that
the effect of fermions on sphaleron heating can lead to new dynamical mechanisms for ending exponential expansion, i.e., a graceful exit for inflation,
if sphaleron heating is disabled during the course of warm inflation.  
They can in principle also switch on the dissipation coefficient $\gamma_{\text{s}}$ dynamically.

Which of these options is realized depends on the time evolution of $\tp$ and $T$, which is model-dependent. For instance, for the type of potentials considered in \eqref{examplaryPotential}, the relation \eqref{temperature} implies that $T$ decreases as a function of $\tp$ if $\beta < 4/3 $ and increases with $\tp$ for $\beta > 4/3 $. Since $\tp$ decreases with time,\footnote{We assume $\tp > 0$ throughout.} this e.g.~implies that chiral masses can switch on or off sphaleron heating if $\beta > 4/3 $ or $\beta < 4/3 $, respectively. 
The same is true for an external chemical potential as discussed in section \ref{ssec:chiral}, with the additional feature that this chemical potential can evolve dynamically if other degrees of freedom are out of equilibrium. This may open up the possibility for a common dynamical origin of the baryon asymmetry and the end of inflation. 
Finally, \eqs \eqref{speciesHubble} and \eqref{speciesHubbleEv} show that the number  of spectator fermions that is needed to re-activate sphaleron heating by Hubble expansion alone scales as 
$N_f \propto \tp^{-2(1+\beta)/7}$, 
i.e., a fixed number $N_f$ of fermions that is present in the theory may be sufficient to enable sphaleron heating at some initial time but fails to do so once $\tp$ drops below a critical value that depends on $f$, $N_c$, $\alpha$, $\beta$ and $g_*$, thereby ending warm inflation.

\section{Conclusions}

Sphaleron heating is a non-perturbative effect that can 
generate a sizeable dissipation term for a slowly evolving pseudoscalar field (axion) in a plasma of non-Abelian gauge bosons without inducing large corrections to its effective potential. This makes it an attractive mechanism to drive warm inflation or warm dark energy. However, it has been pointed out that the mechanism is very sensitive to the presence of light fermions that are charged under the non-Abelian gauge group. If one considers a system that is only composed of the axion, gauge fields and massless fermions, the dissipation term in the axion equation of motion is exactly cancelled due to a contribution arising from the chemical potential generated in the fermion fields. Since all known fermions are massless in the symmetric phase of the electroweak gauge theory, this raises the question how feasible the sphaleron heating mechanism is in the real world. 

In the present work, we investigated how this conclusion changes if one considers a particle physics framework that qualitatively resembles what we know from the SM, such as the presence of Yukawa interactions, other sources of parity violation in addition to the axion coupling or a product gauge group. We find that the presence of any of these ingredients can alleviate the cancellation that disables the sphaleron heating mechanism. In particular, if there is a chirality-violating interaction in any part of the theory that communicates with the sector driving sphaleron heating, the cancellation is incomplete and an effective dissipation rate remains. Even the dilution of the chiral chemical potentials by Hubble expansion alone is capable of enabeling sphaleron heating in the presence of massless fermions for a sufficiently large number of fields. 

Hence, the conclusion that massless fermions disable the sphaleron heating mechanism is far from being generic -- while it is true that the aforementioned cancellation can happen in specific setups, it is easily possible to avoid it by using rather generic ingredients that are known to be realised in Nature and parameters that qualitatively resemble what we see in the SM. This implies that sphaleron heating driven by a dark sector represents a realistic possibility to cause accelerated cosmic expansion, may it be in the early Universe (warm inflation) or now (dark energy radiation). Moreover, additional implications can arise for the production of axionic Dark Matter. We emphasize that our results only show that sphaleron heating in the presence of fermions is feasible in principle and at the background level. The investigation of phenomenological consequences (such as the CMB power spectrum) is model dependent and should be addressed in dedicated works for particular scenarios.
As an additional bonus, some of the mechanisms that enable or disable sphaleron heating depend on the temperature and the value of the axion field. Therefore, their efficiency changes in the course of time evolution, implying that they can be switched on or off dynamically. This offers a novel way to end inflation that is not directly related to the shape of the potential.

\section*{Acknowledgements}
We are grateful to Mehrdad Mirbabayi for triggering the initial idea behind this project, to Valerie Domcke for the inspiration for section \ref{ssec:chiral}, and to Mikhail Shaposhnikov for feedback on the manuscript. This work was supported by the Fonds de la Recherche Scientifique -- FNRS.

\bibliographystyle{JHEP}
\bibliography{references}{}

\end{document}